# Study of non-linear viscoelastic behavior of the human red blood cell

H. Castellini[a], B. Riquelme[b,c]

[a] Departamento de Física, FCEIA (UNR), Pellegrini 250, 2000 Rosario, Argentina;
hcaste@fceia.unr.edu.ar
[b] Área Física, FCByF (UNR), Suipacha 535, 2000 Rosario, Argentina;
briquel@fbioyf.unr.edu.ar
[c] Óptica Aplicada a la Biología, IFIR (CONICET-UNR), Bv. 27 de febrero 210 bis, 2000 Rosario, Argentina; riquelme@ifir-conicet.gov.ar

**Abstract**

The non-linear behavior of human erythrocytes subjected to shear stress was analyzed using data series from the Erythrocyte Rheometer and a theoretical model was developed. Linear behavior was eliminated by means of a slot filter and a sixth order Savisky-Golay filter was applied to the resulting time series that allows the elimination of any possible white noise in the data. A fast Fourier transform was performed on the processed data, which resulted in a series of frequency dominant peaks. Results suggest the presence of a non-linear quadratic term in the Kelvin-Voigt phenomenological model. The correlation dimension studied through recurrence quantification analysis gave $C_2=(2.58\pm0.08)$. Results suggest that the underlying dynamics is the same in each RBC sample corresponding to healthy donors.

## 1. Introduction

In most biomechanical studies, soft tissues have generally been characterized by linear models of viscoelastic materials because they show simultaneous characteristics of elastic solids and viscous fluids. Particularly, the viscoelastic behavior of the red blood cell membrane (RBC), both in human and other animal species has been widely studied [1,2]. RBC viscoelasticity simultaneously involves the storage and dissipation of energy under the application of a force. At the molecular level, the partial dissipation of energy (relaxation phenomenon) is associated with various types of movements of the protein structures, which manifest themselves through important variations of the macroscopic properties.

Viscoelastic materials are usually modelled by consistent approaches such as the Maxwell and Kelvin–Voigt rheological models. The model of quasilinear viscoelasticity was initially proposed by Fung to be applied in soft tissue modelling and takes into account both nonlinearity and time dependence [3]. However, these traditional models were considered as a theoretical approach, since they merely present a partial or phenomenological description of soft tissue. Thus, the original models have been modified by adding more elements in order to achieve a better characterization of the mechanical behavior. Consequently, the complexity of the resulting model was greatly increased due to the presence of these additional terms. Recently, the fractional calculation, an old concept with novel application, has been used in modelling viscoelastic material.

The Fractional Rheological Technique uses differential and integral operators of arbitrary order and it is a powerful tool to study the viscoelastic behavior of materials. This technique has been successful to describe plastic materials and allows a better description of viscoelasticity than traditional models without the introduction of extra physical parameters. Also, there is a series of studies aiming at the characterization of the mechanical effect on soft biological tissues using fractional viscoelastic models.

In this work, we propose a non-linear fractional viscoelastic model to describe the rheological behavior of RBC under stationary and oscillating shear stress. The proposed model take into account the peculiar effects observed in the stationary and dynamical assay using a "Reómetro Eritrocitario", an instrument based on laser diffractometry technique recently developed in our laboratory. The recurrence quantification analysis (RQA) applied to the residual dynamics is aims at corroborating the nonlinear behavior observed in the oscillating shear stress tests.

## 2. Materials and Methods

A theoretical model that fits the observations was developed. The Kelvin-Voigt model was modified by the addition of a non-linear quadratic term associated with the presence of internal normal stress. Also, the ordinary derivative was replaced by a fractional derivative.

*2.1. Biological Samples*

Human blood samples (n=4) from healthy donors were drawn from the antecubital vein, anticoagulated with EDTA/Na2, stored at 4ºC and analyzed within 2 hours after the extraction time, as recommended in the "New guidelines for hemorheological laboratory techniques" [4]. To carry out measurements, 100 µL of each blood sample was poured in 4.5 mL of a solution of polyvinyl pyrrolidone (Sigma PVP360) at 5% (w/v) in PBS (viscosity = (22.0 ± 0.5) cp, pH = (7.40 ± 0.05) and osmolality = (295 ± 8) mOsmol/kg at (25.0 ± 0.5) ºC).

*2.2. Equipment and Measurements*

Data were obtained using the Erythrocyte Rheometer [5], a new instrument developed in our laboratory that give stationary and dynamic viscoelastic parameters of RBCs. Like the first prototype called Erythrodeformeter [6,7,8], the Erythrocyte Rheometer is based on laser diffractometry technique (ektacytometry). This instrument was used to determine the stationary and dynamic viscoelastic parameters of RBCs. In the stationary mode the lower disk rotates at constant speed, whereas in the oscillating mode the lower disk moves at sinusoidal oscillating speeds with frequencies of 0.5, 1 and 1.5 Hz.

2.3 *Mathematical background*

The fractional derivative dates back to the time of the invention of calculus [9] and it was introduced again by Watson in 2004 [10]. The notation $D^u f$ denotes the $u^{th}$ order fractional derivative of the function f, where u > 0. There are many definitions of the fractional derivative: herein, the Caputo definition was used:

$$D^\alpha f(x) = \frac{1}{\Gamma(m-\alpha)} \int_0^t \frac{f^{(m)}(\tau)}{(t-\tau)^{\alpha+1-m}} d\tau \qquad (1)$$

when m – 1 < α < m is the integer , and:

$$D^\alpha f(t) = \frac{d^m f(t)}{dt^m} \qquad (2)$$

When $\alpha = m$. If the periodic function is considered, then the fractional derivative is

$D^\alpha f(t)$, and its frequency response will be:

$$F(D^\alpha f(t))(\omega) = (i\omega)^\alpha F(f(t))(\omega) \qquad (3)$$

where F(f(t))(ω) is the Fourier Transform of f(t).

*2.4 A non-linear fractional viscoelastic model of RBC*

In the present work, the Kelvin-Voigt model was modified by the addition of a non-linear term together with the modification of viscous behavior of RBC by the use of the fractional derivative. The modified Kelvin-Voigt model is:

$$\eta \tau^{\alpha-1} D^\alpha \gamma + \mu \gamma + \epsilon \gamma^2 = \sigma \qquad (4)$$

where σ is the shear stress, γ is the angular deformation, τ is the relax time, η is the viscous coefficient, μ is the Young modulus and ε is slight non-linear behavior.

**3 Processing and results**

*3.1 Dynamic study*

Fig. 1 shows an example of the time series representative of the RBC temporal deformations obtained using the Reómetro Eritrocitario in oscillating mode at 1.5 Hz. The linear behavior was eliminated by means of a slot filter and a sixth-order Savisky-Golay filter was applied to reduce the noise in time series. Then, the data were normalized to null mean and unit variance. A sample space consisting of the data corresponding to RBCs from 4 healthy donors was studied. Fig. 2 shows an example of the curves obtained.
A fast Fourier transform (FFT) was performed on the processed data, leading to a series of frequency dominant peaks (see Fig. 3). These results suggest the presence of a non-linear quadratic term in the Kelvin-Voigt phenomenological model. The peak corresponding to second harmonic can be explained from the rheological equation (4) considering the solution:

$$\gamma(t) = \sum_{n=1}^{\infty} \gamma_n e^{in\omega t} \qquad (5)$$

where, without loss, it can be generally assumed *γ₀=0*. Then:

$$D^\alpha e^{in\omega t} = (in\omega)^\alpha e^{in\omega t} \qquad (6)$$

Replacing equations (5) and (6) in (4) and grouping terms into potentials of $e^{in\omega t}$ results in:

$$\gamma(t)=\gamma_1 e^{i\omega t}+\epsilon\frac{\tau\gamma_1^2}{1+(2i\omega\tau)^\alpha}e^{2i\omega t}+o(\epsilon^2) \quad (7)$$

where the linear solution ($\gamma_1$) and the second harmonic term ($\gamma_2$) are:

$$\gamma_1=\frac{\tau^\alpha\sigma_0/\eta}{1+(iw\tau)^\alpha}\ ; \qquad \gamma_2=\epsilon\frac{\tau\gamma_1^2}{1+(2i\omega\tau)^\alpha} \quad (8)$$

In order to corroborate this hypothesis, the RQA method [11] was used to calculate the Grassberger-Procaccia dimension which corresponds to log-log slope as can be seen in Fig. 4. The value found of the correlation dimension was $C_2$=(2.58±0.08), which is consistent with the oscillating dynamics of a second-order system.

It is very difficult to numerically calculate $\varepsilon$ using only the power spectrum. Yet, it is possible to assume that its value is low so as to be neglected for the stationary study. The numerical calculation of $\varepsilon$ will be considered in a future work.

*3.2 Stationary study*

Fig. 5 shows an example of curves obtained in human RBC under stationary shear stress. The data correspond to longitudinal and transversal axes of the diffraction pattern. This figure shows the presence of overshooting that can be modelled by the fractional derivative. Then, the solution from equation (4) when $\varepsilon=0$ and $\sigma(t)=\sigma_0 H(t-t_0)$ is:

$$\gamma(t)=A_0\left(1-E_\alpha(-(t-t_0)^\alpha/\tau^\alpha)\right) \quad (9)$$

where $E_\alpha(t)$ is the Mittag-Leffler exponential, $A_0$ is a constant and $H(t-t_0)$ is Heaviside function.

Fig. 6 shows the parametric time evolution of $E_\alpha(t)$ for a selected $\alpha$ set (0.5, 1.0, 1.2 and 1.7). The overshooting seen in the curve for stationary shear stress is consistent with $\alpha >1$. To fit this solution to a sample numeric stationary assay, genetic algorithm was used to find $A$, $\alpha$, $t_0$ and $\tau$. Fig. 7 compares a standard solution of Kelvin-Voigt model with the propose model, clearly showing that fitting solution in the experimental data with fractional derivative is better than with the ordinary derivative.

## 4. Discussion

The appearance of quadratic terms in the Kelvin-Voigt equation is associated with the presence of normal stresses against simple shear, which would suggest a limit in the isometric deformation of the RBC.

The use of fractional derivative proves to be a better theoretical framework to characterize human red blood cell behavior under stationary shear stress, without introducing extra parameters such as the use of second-order derivatives. The choice of the nonlinear model is in agreement what is observed in the power spectrum and with the slope observed in the RQA. All this evidence suggests that the behavior of the RBCs would not be that of a linear viscoelastic material.

These results provide information on the non-linear viscoelastic behavior of the erythrocyte membrane. Moreover, this type of analysis will allow to decide on the validity of the application of this same mathematical model in the case of altered RBCs due to hemorheological pathologies (diabetes, hypertension, anemia, parasitosis, etc.) as well *in vitro* alterations caused by different agents (phytochemicals, anesthetics, drugs, etc.).

## Acknowledgment

Authors thank the Universidad Nacional de Rosario for financial support. We would like to thank the staff from the English Department (Facultad de Ciencias Bioquímicas y Farmacéuticas, UNR) for the language correction of the manuscript.

**Figures**

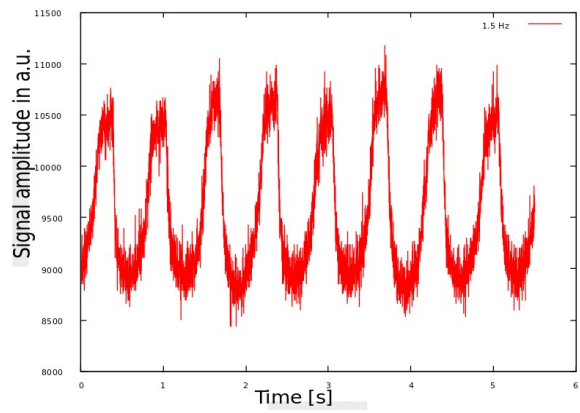

Figure 1: Data at 1.5 Hz of frequency corresponding at major axis of the diffraction pattern recording as a function of time.

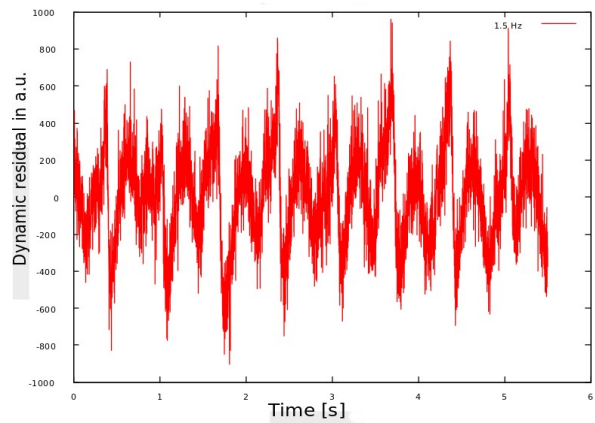

Figure 2: Graphic of results obtained from data processing.

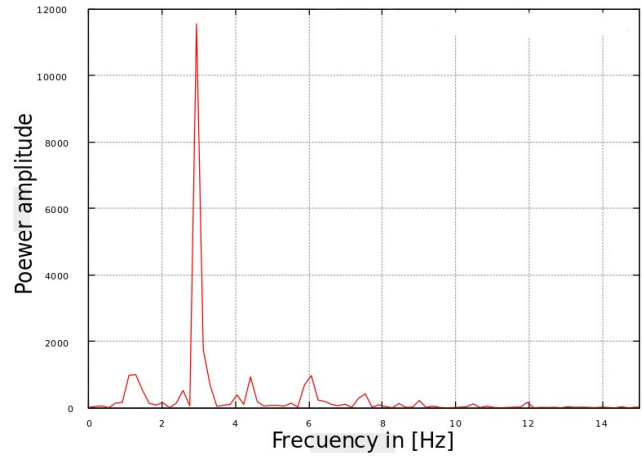

Figure 3: Power spectrum where a peak is seen at twice the excitation frequency.

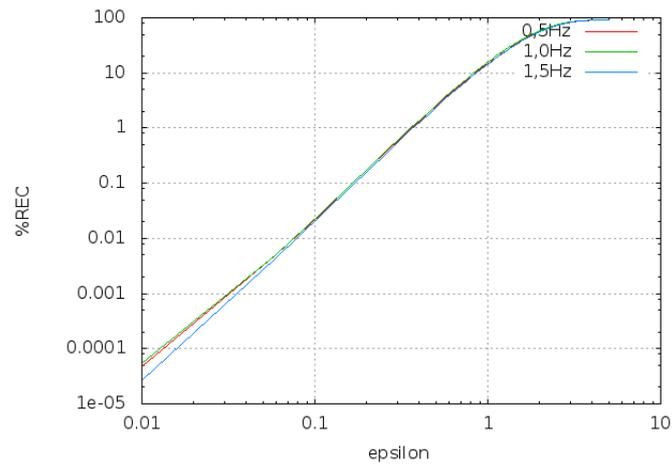

Figure 4: Graphs showing that the residual dynamics is universal of correlation dimension $C_2 = (2.58 \pm 0.08)$.

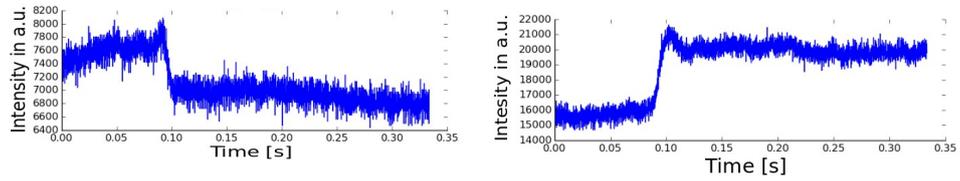

Figure 5: Curves of erythrocyte response to the stationary shear stress corresponding to (a) longitudinal and (b) transversal exes of the diffraction pattern.

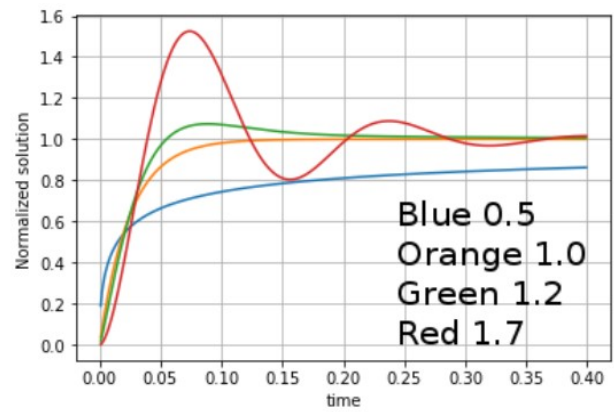

Figure 6: Graph of normalized fitting solution with fractional derivative showing parametric time evolution of $E_\alpha(t)$ for a selected $\alpha$ sets corresponding to 0.5, 1.0, 1.2 and 1.7. The overshooting seen on the curve for stationary shear stress is consistent with $\alpha > 1$.

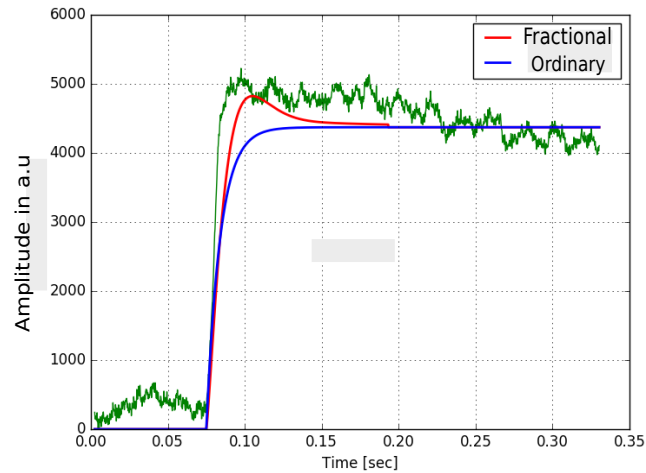

Figure 7: Standard solution of Kelvin-Voigt model (blue line) compared with the propose model (red line). A better fit in the experimental data was obtained using propose model.